\documentclass[12pt]{article}
\usepackage{amsmath}
\usepackage{amssymb}
\input amssym.def
\input amssym

\newcommand{\be}{\begin{equation}}
\newcommand{\ee}{\end{equation}}

\title{Zeeman-type dragging in the Kerr--Newman and NUT
spacetimes}
\author{Nikolai V. Mitskievich\thanks{Physics Department, CUCEI,
University of Guadalajara, Guadalajara, Jalisco, Mexico.}
\thanks{Postal address: Apartado Postal 1-2011, C.P. 44100,
Guadalajara, Jalisco, M\'exico. E-mail:
mitskievich03@yahoo.com.mx} and Luis I. L\'opez
Ben{\'\i}tez\thanks{Mathematics and Physics Department, Instituto
Tecnol\'ogico de Estudios Superiores de Occidente A.C.,
Perif\'erico Sur Manuel G\'omez Mar{\'\i}n 8585, Tlaquepaque,
Jal., C.P. 45090, M\'exico.}}
\date{~}
\begin{document}

\maketitle
\begin{abstract}
In this communication we discuss two distinct Zeeman-type
gravitomagnetic effects deserving attention since they can be
easily characterized in their exact form, not via approximation
procedures. Some observations are also made on gravitoelectric
effects.
\end{abstract}

Gravitoelectromagnetism is an important part of general
relativity, and it is frequently characterized as dragging of
local inertial frames when there is a motion of sources in
Einstein's equations which cannot be globally compensated by any
choice of a non-inertial frame co-moving with these sources (thus
this is related to rotation and/or luminal motion of sources). The
corresponding spacetime is usually stationary, but non-stationary
cases should also lead to gravitoelectromagnetic effects. There is
a vast literature on these subjects, see Refs. 1, 3--12, 17. An
especially interesting aspect is discussed in Refs. 4, 9, first
considered by B. DeWitt and related to an interplay of gravitation
and electromagnetism in conductors and superconductors, i.e.
dragging of electromagnetic field (the usually studied cases of
gravitoelectric and gravitomagnetic effects are of general
relativistic mechanical and time-involving nature).

Let us first consider the case of circular motion of a neutral
test particle in the equatorial plane of the Kerr--Newman
spacetime, $ds^2=$ $\frac{\Delta-a^2\sin^2\vartheta}{\Sigma}dt^2-
\frac{\Sigma}{\Delta}dr^2-\Sigma d\vartheta^2-\frac{1}{\Sigma}
\left[\left(r^2+a^2\right)^2-a^2\Delta\sin^2\vartheta\right]
\sin^2\vartheta d\phi^2 -2a\frac{r^2+a^2+\Delta}{\Sigma}\sin^2
\vartheta dtd\phi$, where $\Sigma=r^2+a^2\cos^2\vartheta$,
$\Delta=r^2+a^2-2Mr+Q^2$ (the Boyer--Lindquist coordinates), while
$\vartheta=\pi/2$. Since the Killing vectors are $\xi^{
\textnormal{I}}=\partial_t$ and $\xi^{ \textnormal{II}}=
\partial_\phi$, there are two conservation laws (for energy and
angular momentum around the $z$ axis; we are working outside the
ergosphere), and the time and azimuthal angle coordinates are
determined unambiguously which gives them an objective meaning.
However we do not even need the corresponding two constants of
motion to be evaluated in this case of the dragging effect: it is
sufficient to consider the $r$-component of the geodesic equation,
$\frac{du_r}{ds}=\frac{1}{2}g_{\alpha\beta, r}u^\alpha u^\beta=0$
($dr/ds=0$ on a circular orbit), thus $g_{tt,r}\dot{t}^2+g_{\phi
\phi,r}\dot{\phi}^2+2g_{t\phi,r}\dot{t} \dot{\phi}=0$, $\dot{F}=
\frac{dF}{ds}$, which reads in our case as $\omega^2-2aS\omega-
S=0$ where $S=\frac{Mr-Q^2}{r^4- a^2Mr+a^2Q^2}$, $\omega=\frac{d
\phi}{dt}$.

There are two roots, $\omega_\pm=-\frac{1}{a\left(1\pm\sqrt{1+
1/(a^2S)} \right)}$, or in terms of the revolution period, $T_\pm
=2\pi\left(\frac{r^2}{\sqrt{Mr-Q^2}}\pm a\right),$ where the first
term describes the ``Newtonian'' revolution period of a test
neutral particle (the coefficients are exactly the same) and the
second one, the dragging effect due to rotation of the central
body. An analogous conclusion was drawn\cite{Mitskie83,
Mitskie05,MitsPul} for motion of a test mass along a circular
equatorial orbit in the Kerr field. We see that in the
Kerr--Newman spacetime the result differs merely in the
``Newtonian'' term which now contains both the mass $M$ and
electric charge $Q$ of the central body, while dragging depends
only on the Kerr parameter $a$ and is exactly the same as in the
Kerr spacetime case (the results are exact and not approximate
ones). This effect is closely related to the Zeeman effect
(spin-orbital interaction).

The second effect occurs in the Taub--NUT spacetime. While the
gravitational mass may be called gravitoelectric charge, the NUT
parameter $l$ is similar (to certain extent) to gravitomagnetic
monopole charge (from the structure of Weyl's tensor the
differences are fairly obvious). The vacuum Taub--NUT metric is
$ds^2=\frac{\Delta}{\Sigma}(dt+2l\cos\vartheta d\phi)^2-
\frac{\Sigma}{\Delta}dr^2-\Sigma\left(d\vartheta^2+ \sin^2
\vartheta d\phi^2\right)$, where $\Delta(r)= r^2-2Mr-l^2$,
$\Sigma(r)=r^2+l^2$; see for more details Refs. 2, 16.

It is clear that there should be an analogue of another case of
electromagnetic Zeeman-type effect (motion of an electrically
charged point-like mass around a centre possessing mass as well as
electric and magnetic monopole charges) if we consider a circular
motion of a (neutral) test mass about the Taub--NUT centre; like
in the electromagnetic case, the orbit has to be centred on the
$z$ axis and not on the origin (central mass).

Then we have to use the conditions $dr=0=d\vartheta$, thus $r$-
and $\vartheta$-components of the geodesic equation, $\frac{d}{ds}
\left(g_{\mu\nu}\frac{dx^\nu}{ds}\right)=\frac{1}{2}g_{\alpha\beta
,\mu}\frac{dx^\alpha}{ds}\frac{dx^\beta}{ds}$, yield $$
\label{NUT} \tan \vartheta= \pm\frac{1}{2l}\sqrt{\frac{g_{tt,r}}{
2r}}\left( \frac{\Sigma^2}{\Delta}-8l^2\right) $$ where $g_{tt,r}
=2\frac{Mr^2+ 2l^2r-Ml^2}{\Sigma^2}$. When $l=0$, the orbit is
centred on the origin ($\tan\vartheta=\infty$), but in the
Taub--NUT case proper, it lies above or under the origin depending
on the relative sign of $l$ and the test particle's angular
momentum, as one can see from the last relation plus an elementary
consideration of two conservation laws (those of energy $\cal E$
and angular momentum $\cal L$, both taken per unit rest mass of
the test particle). Another form of $\vartheta$ then reads
$\cos\vartheta= -\frac{2l{\cal E}}{{\cal L}}$
\cite{Mitskie83,Mitskie05,MitsLB}.

Moreover, in Ref. 10 there was considered the energy (inertial
mass) distribution in the Reissner--Nordstr\"om field, and it was
strictly shown that the electric part of the gravitating mass
density is precisely twice that of the respective inertial one
(electric energy). This point was treated there in terms of
gravitoelectric concepts. Let us recall the Sommerfeld--Lenz
approach\cite{Sommerfeld} discussed from diametrically opposite
viewpoints\cite{Schiff, Rindler}, but now practically forgotten,
primarily, since this approach during decades worked merely in an
intuitive ``deduction'' only of one --- Schwarzschild's ---
solution. However it was later shown\cite{VlaMiHor} that it works
astoundingly well in such a deduction of the
Reissner--Nordstr\"om, Kerr and Kerr--Newman solutions too, so
that all famous eternal black holes can be intuitively reached in
this elementary way (nobody can clearly tell, for what reason).
Here it is only worth mentioning that for charged solutions this
approach needs {\it doubling} the electromagnetic energy
density,\cite{MitsLB,VlaMiHor} precisely in the sense mentioned in
the beginning of this paragraph.

Finally, we should emphasize that, in a contrast to the
Sommerfeld--Lenz approach, gravitoelectromagnetism is not a
hypothesis but a strict consequence of Einstein's gravitation
theory. It even is a paraphrase for a significant part of the
gravitation theory inside the general relativity, the latter
having to be the whole physics under the assumption that spacetime
curvature is included in this picture of universe. Similarly, the
special relativity is not simply a theory of rapid motion but also
is the whole physics under the assumption of properly dealing with
relativistic objects such as any kind of electromagnetic field: in
particular the static Coulomb field is intrinsically relativistic
since the spatial part of its stress-energy tensor is endowed with
the same worth as the temporal-temporal component of this same
tensor. Thus the problem is not so much to verify the theory from
the experimental viewpoint but to refine the experimental means in
physics up to this new level. We are studying general problems of
general relativity to the end of better understanding this theory;
its most exotic features clearly and vividly show its profound
implications, its boundaries, and critical regions of growth of
our knowledge.

\end{document}